\documentclass[aps,prd,reprint,nofootinbib,superscriptaddress]{revtex4-2}

\usepackage{amsmath,amssymb,bm}
\usepackage{mathrsfs}
\allowdisplaybreaks

\newcommand{\Lie}{\mathscr{L}}
\newcommand{\eom}{\approx}
\newcommand{\dd}{\mathrm{d}}
\newcommand{\half}{\tfrac12}

\usepackage[dvipsnames]{xcolor}
\usepackage{hyperref}
\hypersetup{colorlinks=true, linkcolor=Blue, citecolor=Blue, filecolor=Blue, urlcolor=Blue}
\usepackage[nameinlink]{cleveref}
\crefname{equation}{Eq.}{Eqs.}
\Crefname{equation}{Equation}{Equations}
\crefname{section}{Sec.}{Secs.}
\Crefname{section}{Sec.}{Secs.}
\crefname{table}{Table}{Tables}
\Crefname{table}{Table}{Tables}

\begin{document}

\title{Bessel-Hagen currents for linearised Weyl-squared gravity}

\author{Michael \surname{Hobson}}
\email{mph@mrao.cam.ac.uk}
\affiliation{Astrophysics Group, Cavendish Laboratory, J.J. Thomson Avenue, Cambridge, CB3 0HE, UK}

\author{Will Barker}
\email{barker@fzu.cz}
\affiliation{Institute of Physics of the Czech Academy of Sciences, Na Slovance 1999/2, 182 00 Prague 8, Czechia}

\author{Anthony \surname{Lasenby}}
\email{a.n.lasenby@mrao.cam.ac.uk}
\affiliation{Astrophysics Group, Cavendish Laboratory, J.J. Thomson Avenue, Cambridge, CB3 0HE, UK}
\affiliation{Kavli Institute for Cosmology, Madingley Road, Cambridge, CB3 0HA, UK}

\date{\today}

\begin{abstract}
We consider the Bessel-Hagen construction for linearised Weyl-squared gravity,
whose action depends on the metric perturbation~$h_{\mu\nu}$ through the
linearised Weyl tensor~$C^{(1)}_{\mu\nu\rho\sigma}$. This provides a curvature-based
comparison with the Fierz-Pauli theory, in which the action contains first
derivatives of~$h_{\mu\nu}$ whereas the first local gauge-invariant field
strength contains two, and the construction determines a gauge-invariant class
of Noether currents without selecting a preferred gauge-invariant local
energy-momentum tensor. Since the Weyl tensor is the irreducible spin-2 part
of the curvature, it is of particular interest to ask whether an action built
from it admits a local representative analogous to the Maxwell tensor. We show
that the Bessel-Hagen construction extends to the conformal group, acting on
the fields on a fixed Minkowski background, and yields a gauge-invariant
conserved-current class for each generator. Nonetheless, no nonzero local,
polynomial, symmetric rank-two tensor of mass dimension four, quadratic in
$h_{\mu\nu}$ and with constant dimensionless coefficients and no explicit
coordinate dependence, is both strictly invariant under the linearised
diffeomorphism and Weyl gauge symmetries and conserved on the Bach shell.
The gauge symmetries restrict such a tensor to depend on~$C^{(1)}$ and its
derivatives, while the dimension and field degree permit only terms quadratic
in~$C^{(1)}$ without further derivatives. Four-dimensional Weyl identities then
reduce all candidates to pure traces, whose conservation requires their
coefficients to vanish. Thus, building the action from a gauge-invariant
curvature does not recover the electromagnetic result. A first-order
gauge-covariant field strength may provide a route to such a representative,
although its presence alone is not sufficient.

\end{abstract}

\maketitle

\section{Introduction}

The localisation of gravitational field energy has long presented difficulties
in general relativity. In normal coordinates the connection coefficients
vanish at a chosen point, although the tidal curvature need not, so that an
energy density constructed from first derivatives of the metric can be made
to vanish there. For the same reason, the object obtained from Noether's
theorem for a rigid translation, which is built from first derivatives, is not
a tensor. These observations explain the
familiar difficulty with such constructions, but do not exclude every possible
pointwise local expression. The classical energy-momentum
pseudo-tensors~\cite{Einstein1916,LandauLifshitz,BergmannThomson1953,Moller1958}
are conserved and recover the accepted total energy of an isolated system
under suitable asymptotic conditions and coordinate choices, although they
depend on the coordinates or frame and are not unique.

One approach to this non-uniqueness is to retain a local density while fixing
the gauge. Examples include the tensor of Butcher, Hobson and
Lasenby~\cite{ButcherHobsonLasenby2010,ButcherLasenbyHobson2012,ButcherHobsonLasenby2012},
introduced in harmonic gauge and subsequently generalised to other gauges,
and the Fierz-tensor formulation of T\'oth~\cite{Toth2022}, which uses first
derivatives of~$h_{\mu\nu}$ and satisfies the dominant energy condition in a
class of gauges containing the transverse-traceless gauge. T\'oth has
recently extended this construction to the dilatation current of linearised
gravity, an ordinary Noether symmetry, and, through a generalised Noether
theorem that incorporates gauge conditions, to special conformal currents in
generalised harmonic gauges~\cite{Toth2026}. Alternatively, one
may consider asymptotic or quasi-local charges. The former are well defined
under the appropriate asymptotic conditions, whereas the latter retain choices
of boundary and reference structure~\cite{ADM1962,Szabados2009}. For a theory
linearised about Minkowski spacetime, the coordinate freedom is expressed as
the gauge redundancy of a spin-2 field, and one may ask whether the conserved
Noether current associated with a rigid symmetry admits a strictly
gauge-invariant local representative, or whether gauge invariance holds only
for its equivalence class. We address this question by comparison with
electromagnetism.

For electromagnetism (EM) in four-dimensional Minkowski spacetime, the
Bessel-Hagen method~\cite{BesselHagen1921} gives the standard gauge-invariant
energy-momentum tensor directly from Noether's theorem. Since Maxwell theory
is conformally invariant, the method applies to every conformal Killing
vector~$\zeta^\mu$. The variation of the four-potential is supplemented by a
gauge transformation~$A_\mu\to A_\mu+\partial_\mu\alpha$, with
$\alpha=A_\nu\zeta^\nu$, so that the combined form variation is itself gauge
invariant, $\delta_0 A_\mu=\zeta^\nu F_{\mu\nu}$. The corresponding current is
$J^\mu=-\zeta^\nu\tau^\mu{}_\nu$, where~$\tau^\mu{}_\nu$ is the Maxwell tensor;
hence all the conformal currents are coordinate-weighted contractions of the
same gauge-invariant tensor~\cite{HobsonLasenbyBarker2024}. In this construction,
the gauge-invariant field strength~$F_{\mu\nu}\sim\partial A$ contains only
first derivatives of the potential.

The analogous construction for the massless spin-2 field described by the
Fierz-Pauli action~\cite{FierzPauli1939} has a different
outcome~\cite{HobsonBarkerLasenbyFP}. This action is invariant under the
Poincar\'e group and under dilations, but not under special conformal
transformations~\cite{Toth2026}, and the spacetime generators considered in
our Fierz-Pauli analysis~\cite{HobsonBarkerLasenbyFP} were restricted to the
Poincar\'e group. Although the action is invariant under
$h_{\mu\nu}\mapsto h_{\mu\nu}+\partial_\mu\xi_\nu+\partial_\nu\xi_\mu$,
it is constructed from first derivatives of~$h_{\mu\nu}$, whereas the first
local gauge-invariant field strength is the linearised curvature,
$R^{(1)}\sim\partial\partial h$. Consequently, the Bessel-Hagen form variation
has the structure of a connection, rather than a curvature, and one obtains a
gauge-invariant equivalence class of currents modulo divergences of
antisymmetric superpotentials and terms proportional to the field equations.
This equivalence is understood in the sense of the characteristic cohomology
of Barnich, Brandt and Henneaux~\cite{BarnichBrandtHenneaux2000}; the associated
non-uniqueness in linearised gravity has been studied in
detail~\cite{BakerKK2021,TaylorBaker2024}, and the construction does not select a
preferred gauge-invariant local tensor~\cite{DeserHenneaux1995,MagnanoSokolowski2002}.

It is therefore of interest to consider whether an action constructed directly
from the linearised curvature admits a local energy-momentum tensor analogous
to that of electromagnetism. We consider linearised Weyl-squared gravity, which
is obtained as the quadratic flat-space limit of conformal (Weyl) gravity,
\begin{equation}
 S_{\rm W}=\alpha\int\dd^4x\,\sqrt{-g}\,
 C_{\mu\nu\rho\sigma}C^{\mu\nu\rho\sigma},\label{eq:confgrav}
\end{equation}
which has been investigated as an alternative to general relativity, including
as a possible explanation of galactic rotation curves without dark
matter~\cite{MannheimKazanas1989,cgrotcurves}. The quadratic action depends on
$h_{\mu\nu}$ only through~$C^{(1)}_{\mu\nu\rho\sigma}$. Our reason for choosing
this theory is the comparison with the Maxwell action: just as the latter is
quadratic in the irreducible spin-1 field strength~$F_{\mu\nu}$, the Weyl-squared
action is quadratic in the irreducible spin-2 part of the curvature. By
contrast, $R^2$ uses its scalar part, while a general quadratic-curvature action
mixes spins. In four dimensions, Weyl-squared gravity is the unique parity-even,
non-topological quadratic-curvature theory constructed solely from this spin-2
curvature, and is equivalently the parity-even, non-topological conformally
invariant quadratic-curvature theory.\footnote{The parity-odd
$C^{(1)}{}^\ast\!C^{(1)}$ Pontryagin density and the Gauss--Bonnet term are
topological in four dimensions. Their actions are conformally invariant, in
the Gauss--Bonnet case up to boundary terms, although the Euler density is not
in general pointwise Weyl invariant.}

Conformal invariance thus follows from the choice of curvature in this
comparison. In particular, $C^{(1)}$ is invariant under both linearised
diffeomorphisms and Weyl rescalings, and the spacetime symmetries of the action
allow the Bessel-Hagen construction to be extended beyond the Poincar\'e group.
Moreover, as we show below, the Noether momentum is a symmetric projection of
$C^{(1)}$ itself, which suggests a closer correspondence with the electromagnetic
construction than occurs for Fierz-Pauli. A generalised Bessel-Hagen method has
recently been applied by Faria~\cite{Faria2025} to fourth-order conformally
invariant gravitational theories, yielding an explicit energy-momentum
pseudo-tensor. Here we consider whether a strictly gauge-invariant local tensor
can exist for the pure linearised Weyl-squared theory, and determine the
gauge-invariant content of its Noether currents.

The canonical energy of higher-derivative gravity is of Ostrogradski type and
is not positive definite~\cite{Stelle1978}. Our question concerns instead the
existence of a strictly gauge-invariant representative of a Noether current
class, for which positivity is not assumed. The analysis depends on the allowed
local expressions and the four-dimensional curvature identities they obey,
and makes no use of the sign of the canonical energy. Conserved charges
associated with the class remain the relevant gauge-invariant quantities,
subject to the appropriate boundary conditions, as in the background and
asymptotic charge constructions for higher-curvature
gravity~\cite{DeserTekin2002,DeserTekin2003}.\footnote{These difficulties need not
apply to higher-derivative counterterms in a truncated effective theory of
gravity~\cite{Donoghue:1994dn,Donoghue:1993eb}. Such operators arise when
Fierz-Pauli theory is extended by consistent interactions through the
perturbative Noether procedure, but a perturbative treatment of the low-energy
theory, or equivalently an order-reduction scheme, does not introduce
Ostrogradski instabilities. The pure Weyl-squared theory considered here does
not have this interpretation, as is already indicated by the absence of an
ultraviolet scale. Stelle theory~\cite{Stelle1977}, on the other hand, retains
the higher derivatives to obtain renormalisability and consequently contains
ghost modes.}

We first show that each conformal generator acts on the field on a fixed
Minkowski background with a weight contribution~$2\sigma h_{\mu\nu}$, where
$\partial_\mu\zeta_\nu+\partial_\nu\zeta_\mu=2\sigma\eta_{\mu\nu}$, and that the
associated conserved-current class is gauge invariant. In particular, the
second derivatives of~$\zeta^\mu$ occurring for special conformal
transformations can be included in an internal Weyl gauge variation. We then
prove that the translation current has no strictly gauge-invariant local
representative within the class of polynomial, symmetric rank-two tensors of
mass dimension four, quadratic in~$h_{\mu\nu}$, with no explicit coordinate
dependence or external scale.

The paper is arranged as follows. We review the electromagnetic construction
in~\cref{sec:em}, before introducing linearised Weyl-squared theory and its
conformal transformations in~\cref{sec:theory,sec:background}. The current
family and the gauge invariance of its class are established in~\cref{sec:class},
and the non-existence result is proved in~\cref{sec:nogo}. In~\cref{sec:discussion}
we compare this result with Fierz-Pauli and linearised Gauss--Bonnet gravity,
whose proposed curvature-quadratic tensor is the identically vanishing
four-dimensional Lanczos tensor~\cite{BakerKuzmin2019,BakerKuzmin2021}. We also
discuss the possible role of first-order gauge-covariant field strengths,
suggested by electromagnetism and by the manifestly covariant currents of Weyl
gauge theory~\cite{HobsonLasenbyBarker2024}. Companion
papers~\cite{HobsonBarkerLasenbyEMT,HobsonBarkerLasenbyT4} consider a linearised
criterion for the metric spin-2 theories within their scope and a separate
nonlinear internal-source construction, respectively; the Bessel-Hagen
analysis of the dilation sector remains open.

We use the metric signature~$(+\,-\,-\,-)$ and the curvature conventions of
Ref.~\cite{cgrotcurves}:
${R^\rho}_{\sigma\mu\nu}=\partial_\mu{\Gamma^\rho}_{\sigma\nu}
-\partial_\nu{\Gamma^\rho}_{\sigma\mu}
+{\Gamma^\rho}_{\lambda\mu}{\Gamma^\lambda}_{\sigma\nu}
-{\Gamma^\rho}_{\lambda\nu}{\Gamma^\lambda}_{\sigma\mu}$ and
$R_{\mu\nu}={R^\rho}_{\mu\rho\nu}$. Symmetrisation and antisymmetrisation
both carry unit weight,~$A_{(\mu\nu)}=\half(A_{\mu\nu}+A_{\nu\mu})$ and
$A_{[\mu\nu]}=\half(A_{\mu\nu}-A_{\nu\mu})$; with this convention the Weyl tensor
\cref{eq:weyl} below carries its standard four-dimensional coefficients.

\section{Electromagnetic precursor}
\label{sec:em}

We begin with electromagnetism, for which the Bessel-Hagen construction gives
a local gauge-invariant energy-momentum tensor and also determines the
currents associated with conformal transformations. The Maxwell action in
four-dimensional Minkowski spacetime is
\begin{equation}
 S_{\rm EM}=-\tfrac14\int\dd^4x\,F_{\mu\nu}F^{\mu\nu},
 \qquad F_{\mu\nu}=\partial_\mu A_\nu-\partial_\nu A_\mu,
\label{eq:emaction}
\end{equation}
where indices are raised with~$\eta^{\mu\nu}$. We regard~$A_\mu$ as a one-form
field on a fixed Minkowski background and use Cartesian coordinates throughout.
The metric and coordinate measure~$\dd^4x$ are therefore fixed, while symmetries
are realised by active variations of the field. In particular, the action is
invariant under the internal~$U(1)$ transformation
$A_\mu\to A_\mu+\partial_\mu\alpha$.

\subsection{Conformal invariance on the fixed background}

The spacetime transformations relevant to~\cref{eq:emaction} are the conformal
symmetries of Minkowski spacetime, rather than arbitrary diffeomorphisms. Their
generators are conformal Killing vectors (CKVs), which satisfy
\begin{equation}
 \partial_\mu\zeta_\nu+\partial_\nu\zeta_\mu=2\sigma\eta_{\mu\nu},
 \qquad \sigma=\tfrac14\partial_\rho\zeta^\rho,
\label{eq:ckv}
\end{equation}
with general solution
\begin{equation}
 \zeta^\mu=a^\mu+{\omega^\mu}_\nu x^\nu+\rho\, x^\mu
 +c^\mu x^2-2(c\cdot x)x^\mu ,
\label{eq:ckvgen}
\end{equation}
for which~$\sigma = \rho - 2c\cdot x$. Taking the active variation of~$A_\mu$
to be that induced by its one-form character gives
\begin{equation}
 \delta^{(\zeta)}_0 A_\mu=-\Lie_\zeta A_\mu
 =-\zeta^\rho\partial_\rho A_\mu-A_\rho\partial_\mu\zeta^\rho,
\label{eq:emAtransf}
\end{equation}
where~$\Lie_\zeta$ denotes the Lie derivative along the vector field~$\zeta^\mu$,
the field strength transforms as a two-form,
$\delta^{(\zeta)}_0 F_{\mu\nu}=-\Lie_\zeta F_{\mu\nu}$, and the density transforms
as a scalar density of weight one,
\begin{equation}
 \delta^{(\zeta)}_0\mathcal{L}_{\rm EM}
 =-\partial_\mu(\zeta^\mu\mathcal{L}_{\rm EM}),
\label{eq:emLdensity}
\end{equation}
so that~$\delta^{(\zeta)}_0 S_{\rm EM}=0$. No additional scale factor is required
in~\cref{eq:emAtransf}; in the convention of
Refs.~\cite{CGTpaper,eWGTpaper}, the one-form transformation corresponds to a
global Weyl weight~$w(A_\mu)=-1$.\footnote{In this convention a dilation
$x'=e^\rho x$ gives~$A'_\mu(x')=e^{w\rho}A_\mu(x)$, so the inverse-Jacobian
factor of an ordinary one-form is included in the weight~$w=-1$~\cite{eWGTpaper}.
The conformal invariance of this assignment is specific to four dimensions:
under a passive Weyl rescaling~$\gamma_{\mu\nu}\to\Omega^2\gamma_{\mu\nu}$,
the density~$\sqrt{-\gamma}\,\gamma^{\mu\rho}\gamma^{\nu\sigma}
F_{\mu\nu}F_{\rho\sigma}$ acquires a factor~$\Omega^{n-4}$.}
In the active description used here, the contribution that would arise from
the Jacobian of a conformal coordinate transformation appears as the term
$\partial_\mu\zeta^\mu=4\sigma$ in the divergence~\cref{eq:emLdensity}.

\subsection{The Noether momentum is the field strength}

Since the Maxwell Lagrangian depends on~$A_\mu$ only through its first
derivatives, its conjugate momentum is given by
\begin{equation}
 \Pi^{\mu|\sigma}\equiv
 \frac{\partial\mathcal{L}_{\rm EM}}{\partial(\partial_\mu A_\sigma)}
 =-F^{\mu\sigma}.
\label{eq:emmomentum}
\end{equation}
Thus the conjugate momentum is itself gauge invariant. An analogous relation
holds between the Noether momentum and the Weyl tensor in the spin-2 theory
[\cref{eq:momentum}], although the latter contains second derivatives of the
field.

\subsection{Bessel-Hagen current and the Maxwell tensor}

Neither the form variation~\cref{eq:emAtransf} nor the canonical Noether current
obtained from it is gauge invariant. Since the action is also invariant under
internal~$U(1)$ transformations, however, one may follow Bessel-Hagen and
combine the active conformal variation with a gauge variation to obtain
$\delta_0 A_\mu=-\Lie_\zeta A_\mu+\partial_\mu\alpha$. Choosing
$\alpha=A_\nu\zeta^\nu$ then gives
\begin{equation}
 \delta_0 A_\mu=\zeta^\nu F_{\mu\nu},
\label{eq:embhform}
\end{equation}
which depends only on the gauge-invariant field strength. Together with
$\Pi^{\mu|\sigma}=-F^{\mu\sigma}$, this ensures that the Noether current
$J^\mu=\Pi^{\mu|\sigma}\delta_0 A_\sigma+\zeta^\mu\mathcal{L}_{\rm EM}$ is a
gauge-invariant local expression, and one finds
\begin{equation}
\begin{aligned}
 J^\mu &=-\zeta^\nu\,\tau^\mu{}_\nu,\\
 \tau^\mu{}_\nu &=-\Big(F^{\mu\sigma}F_{\nu\sigma}
 -\tfrac14\delta^\mu_\nu F^{\rho\sigma}F_{\rho\sigma}\Big),
\end{aligned}
\label{eq:maxwelltensor}
\end{equation}
where~$\tau^\mu{}_\nu$ is the standard symmetric, traceless Maxwell
energy-momentum tensor, whose divergence vanishes on shell. Since this tensor
is independent of~$\zeta^\mu$, inserting~\cref{eq:ckvgen} expresses every
conformal current as a coordinate-weighted contraction of it:
\begin{equation}
\begin{split}
 J^\mu={}&-a^\nu\tau^\mu{}_\nu
 +\tfrac12\omega^{\rho\sigma}\big(x_\rho\tau^\mu{}_\sigma-x_\sigma\tau^\mu{}_\rho\big)\\
 &-\rho\,x^\nu\tau^\mu{}_\nu
 +c^\nu\big(2x_\nu x^\rho-\delta^\rho_\nu x^2\big)\tau^\mu{}_\rho,
\end{split}
\label{eq:emmoments}
\end{equation}
and the four conservation laws are consequently implied by the properties
\begin{equation}
 \partial_\mu\tau^\mu{}_\nu\eom0,
 \qquad \tau_{[\mu\nu]}=0,
 \qquad \tau^\mu{}_\mu=0 ;
\label{eq:emconslaws}
\end{equation}
the dilation and special conformal conservation laws follow from
tracelessness in addition to conservation and symmetry. In electromagnetism,
the conformal currents can therefore all be expressed in terms of one strictly
gauge-invariant local tensor. A detailed account of this calculation is given
in the appendix of Ref.~\cite{HobsonLasenbyBarker2024}.

\subsection{Comparison with the spin-2 construction}
\label{sec:emwhy}

Two features of the electromagnetic result are useful for the comparison below.
The field strength~$F_{\mu\nu}\sim\partial A$ contains one derivative of the
potential, and its field equation~$\partial_\mu F^{\mu\nu}=0$ contains one
further derivative, allowing the conservation of the quadratic Maxwell tensor.
Moreover, the combination
$F_{\mu\rho}F_\nu{}^\rho-\tfrac14\eta_{\mu\nu}F^{\rho\sigma}F_{\rho\sigma}$
is not identically zero. For linearised Weyl-squared theory the momentum is
again gauge invariant, but the field strength contains two derivatives of
$h_{\mu\nu}$ and the field equation contains two further derivatives. In
addition, a four-dimensional identity makes the corresponding trace-free
quadratic rank-two tensor vanish. We will establish these differences
explicitly before considering their consequences for a local representative.

\section{Linearised Weyl-squared theory}
\label{sec:theory}

For a symmetric field~$h_{\mu\nu}$ on Minkowski spacetime, we define the
linearised Riemann tensor and its Ricci, scalar and Weyl parts by
\begin{equation}
\begin{aligned}
 R^{(1)}_{\mu\nu\rho\sigma}
 &= \half\big(
   \partial_\nu\partial_\rho h_{\mu\sigma}
 + \partial_\mu\partial_\sigma h_{\nu\rho} \\
 &\quad
 - \partial_\nu\partial_\sigma h_{\mu\rho}
 - \partial_\mu\partial_\rho h_{\nu\sigma}\big), \\
 R^{(1)}_{\mu\nu}&=\eta^{\rho\sigma}R^{(1)}_{\rho\mu\sigma\nu}, \\
 R^{(1)}&=\eta^{\mu\nu}R^{(1)}_{\mu\nu}, \\
 C^{(1)}_{\mu\nu\rho\sigma}
 &= R^{(1)}_{\mu\nu\rho\sigma}
 -\big(\eta_{\mu[\rho}R^{(1)}_{\sigma]\nu}-\eta_{\nu[\rho}R^{(1)}_{\sigma]\mu}\big) \\
 &\quad
 +\tfrac13 R^{(1)}\eta_{\mu[\rho}\eta_{\sigma]\nu}.
\end{aligned}
\label{eq:weyl}
\end{equation}
The free conformal spin-2 action is taken to be
\begin{equation}
\begin{aligned}
 S_{\rm C}[h]&=\half\int\dd^4x\,
 C^{(1)}_{\mu\nu\rho\sigma}C^{(1)\mu\nu\rho\sigma}, \\
 \mathcal{L}_{\rm C}&=\half C^{(1)}C^{(1)} .
\end{aligned}
\label{eq:action}
\end{equation}
The action admits the two internal gauge transformations
\begin{equation}
\begin{aligned}
 \delta_\xi h_{\mu\nu}&=D_\xi h_{\mu\nu}
 \equiv\partial_\mu\xi_\nu+\partial_\nu\xi_\mu, \\
 \delta_\omega h_{\mu\nu}&=2\omega\,\eta_{\mu\nu},
\end{aligned}
\label{eq:gauges}
\end{equation}
corresponding to linearised diffeomorphisms and Weyl rescalings, respectively.
The linearised Weyl tensor is invariant under each transformation,
\begin{equation}
 \delta_\xi C^{(1)}_{\mu\nu\rho\sigma}=0,
 \qquad
 \delta_\omega C^{(1)}_{\mu\nu\rho\sigma}=0 .
 \label{eq:Cgaugeinv}
\end{equation}

\subsection{The Noether momentum is the Weyl tensor}

The dependence of~\cref{eq:action} on~$h_{\mu\nu}$ is entirely through its
second derivatives, since~$C^{(1)}\sim\partial\partial h$. Hence the only
non-vanishing conjugate momentum is that associated with these derivatives:
\begin{equation}
\begin{gathered}
 \Pi^{\alpha\beta|\mu\nu}\equiv
 \frac{\partial\mathcal{L}_{\rm C}}{\partial(\partial_\alpha\partial_\beta h_{\mu\nu})}, \\
 \frac{\partial\mathcal{L}_{\rm C}}{\partial h_{\mu\nu}}
 =\frac{\partial\mathcal{L}_{\rm C}}{\partial(\partial_\alpha h_{\mu\nu})}=0 .
\end{gathered}
\label{eq:momdef}
\end{equation}
To evaluate this momentum, one may use the tracelessness of~$C^{(1)}$ to write
$C^{(1)\kappa\lambda\rho\sigma}\delta C^{(1)}_{\kappa\lambda\rho\sigma}
=C^{(1)\kappa\lambda\rho\sigma}\delta R^{(1)}_{\kappa\lambda\rho\sigma}$.
Substituting the variation of~\cref{eq:weyl} and using the curvature symmetries
then yields
\begin{equation}
 \Pi^{\alpha\beta|\mu\nu}=2\,C^{(1)\mu\alpha\beta\nu}
 \label{eq:momentum}
\end{equation}
up to terms antisymmetric in~$(\alpha\beta)$ or~$(\mu\nu)$, which vanish when
contracted with~$\partial_\alpha\partial_\beta h_{\mu\nu}$. More explicitly,
the fully symmetrised representative is
$\Pi^{\alpha\beta|\mu\nu}=\tfrac12\big(
C^{(1)\mu\alpha\beta\nu}+C^{(1)\mu\beta\alpha\nu}
+C^{(1)\nu\alpha\beta\mu}+C^{(1)\nu\beta\alpha\mu}\big)$, since the variable
with respect to which the derivative is taken is symmetric in both index
pairs, whereas~$C^{(1)\mu\alpha\beta\nu}$ is not. The momentum is therefore a
projection of the Weyl tensor and is invariant under both gauge
symmetries~\cref{eq:gauges}. This differs from the Fierz-Pauli momentum,
$\Pi^{\mu|\alpha\beta}\sim\partial h$, which is gauge dependent, and provides
the correspondence with electromagnetism that motivates the present analysis.

\subsection{Symplectic potential, Bach tensor and Noether identities}

For a Lagrangian density containing second derivatives, integration by parts
gives the Ostrogradski variational identity
\begin{equation}
 \delta\mathcal{L}_{\rm C}
 =\mathcal{B}^{\mu\nu}\delta h_{\mu\nu}
 +\partial_\alpha\Theta^\alpha(\delta h),
\label{eq:var}
\end{equation}
where the Euler--Lagrange expression is the linearised Bach tensor and the
symplectic potential contains both the momentum and its derivative:
\begin{equation}
\begin{gathered}
 \mathcal{B}^{\mu\nu}
 =\partial_\alpha\partial_\beta\Pi^{\alpha\beta|\mu\nu}
 =2\,\partial_\alpha\partial_\beta C^{(1)\mu\alpha\beta\nu}, \\
 \Theta^\alpha(\delta h)
 =\Pi^{\alpha\beta|\mu\nu}\partial_\beta\delta h_{\mu\nu}
 -\big(\partial_\beta\Pi^{\alpha\beta|\mu\nu}\big)\delta h_{\mu\nu}.
\end{gathered}
\label{eq:bach}
\end{equation}
The symmetries and tracelessness of~$C^{(1)}$ imply the off-shell Noether
identities
\begin{equation}
 \partial_\mu\mathcal{B}^{\mu\nu}\equiv0,
 \qquad
 \mathcal{B}^\mu{}_\mu\equiv0,
 \label{eq:bachidentities}
\end{equation}
which correspond to the diffeomorphism and Weyl gauge symmetries,
respectively. The equations of motion are~$\mathcal{B}^{\mu\nu}=0$, and the
first identity is analogous to~$\partial_\mu\mathcal{E}^{\mu\nu}\equiv0$ for
the Fierz-Pauli Euler--Lagrange expression.

\section{Conformal transformations and invariance of the action}
\label{sec:background}

We now consider the action of the conformal group on~$h_{\mu\nu}$, using the
CKVs defined in~\cref{eq:ckv,eq:ckvgen}. For dilations and special conformal
transformations, one has~$\sigma=\rho$ and~$\sigma=-2c\cdot x$, respectively.
In the latter case the second derivatives of~$\zeta^\mu$ do not vanish; more
generally, differentiating the conformal Killing equation gives
\begin{equation}
 \partial_\mu\partial_\nu\zeta_\rho
 =\eta_{\mu\rho}\partial_\nu\sigma+\eta_{\nu\rho}\partial_\mu\sigma
 -\eta_{\mu\nu}\partial_\rho\sigma .
 \label{eq:ddckv}
\end{equation}

As in the electromagnetic calculation, the conformal transformations are
realised actively on the field in fixed Cartesian coordinates. They are the
finite-dimensional conformal symmetries of Minkowski spacetime and do not
represent arbitrary diffeomorphisms of the background. With
$\eta_{\mu\nu}=\mathrm{diag}(1,-1,-1,-1)$ and~$\dd^4x$ held fixed, the required
field variation is
\begin{equation}
 \Delta^\zeta_{\mu\nu}\equiv\delta^{(\zeta)}_0 h_{\mu\nu}
 =-\Lie_\zeta h_{\mu\nu}+2\sigma h_{\mu\nu},
\label{eq:Qzeta}
\end{equation}
where the second term supplies the conformal weight of~$h_{\mu\nu}$. In the
convention of Refs.~\cite{CGTpaper,eWGTpaper}, its global Weyl weight is zero,
whereas an ordinary~$(0,2)$ tensor has weight~$-2$.\footnote{For a dilation
$x'=e^\rho x$, the convention~$\Phi'(x')=e^{w\rho}\Phi(x)$ includes the tensor
transformation in~$w$. The two inverse-Jacobian factors of an ordinary~$(0,2)$
tensor give~$w=-2$, while the additional term~$2\sigma h_{\mu\nu}$ raises this
to zero. Equivalently, the weight~$+2$ of the metric under
$g_{\mu\nu}\to\Omega^2g_{\mu\nu}$ offsets the weight~$-2$ from coordinate
transport; see~\cref{eq:phisigma}.}
Modulo the internal gauge variations~$D_\xi h+2\omega\eta$, this is the unique
homogeneous weight contribution to~$-\Lie_\zeta h_{\mu\nu}$ for which the
linearised Weyl tensor transforms as a weight-two conformal primary,
\begin{equation}
 \delta^{(\zeta)}_0 C^{(1)}_{\mu\nu\rho\sigma}
 =-\Lie_\zeta C^{(1)}_{\mu\nu\rho\sigma}+2\sigma C^{(1)}_{\mu\nu\rho\sigma},
\label{eq:Ccovariant}
\end{equation}
including for special conformal transformations.\footnote{The difference
between~$-\Lie_\zeta C^{(1)}[h]$ and~$C^{(1)}[-\Lie_\zeta h]$ consists of terms
containing~$\partial_\mu\partial_\nu\zeta^\rho$, arising because the Lie
derivative does not commute with the flat-space curvature operator. The
$\sigma$-gradient terms in~$C^{(1)}[2\sigma h]$ cancel these contributions by
virtue of~\cref{eq:ddckv}. A related cancellation appears in~\cref{eq:dQ}.}
Equivalently, the Lagrangian transforms as a scalar density of weight one,
\begin{equation}
 \delta^{(\zeta)}_0\mathcal{L}_{\rm C}
 =-\partial_\mu(\zeta^\mu\mathcal{L}_{\rm C}),
\label{eq:Ldensity}
\end{equation}
and hence~$\delta^{(\zeta)}_0 S_{\rm C}=0$. Together with~\cref{eq:Cgaugeinv},
this establishes invariance of the action under the conformal transformations
and under both internal gauge symmetries. The weight term is required for the
former statement: for a dilation~$\zeta^\mu=\lambda x^\mu$, the variation
$\delta_0h_{\mu\nu}=-\Lie_\zeta h_{\mu\nu}$ alone would give
$\delta_0S_{\rm C}=-2\lambda\int C^{(1)2}\,\dd^4x=-4\lambda S_{\rm C}$.

No varying volume element is introduced in this active description. Under a
passive conformal coordinate transformation, one would instead use the
conformally related metric~$\gamma_{\mu\nu}$ and the measure
$\sqrt{-\gamma}\,\dd^4x$. Here the corresponding contributions are accounted
for by the weight term in~\cref{eq:Qzeta} and by
$(\partial_\mu\zeta^\mu)\mathcal{L}_{\rm C}=4\sigma\mathcal{L}_{\rm C}$ in the
divergence~\cref{eq:Ldensity}, which expresses the conformal invariance of the
action with the fixed coordinate measure.

One may also derive~\cref{eq:Qzeta} by linearising a combined diffeomorphism and
Weyl transformation of the full-metric action
$\int\sqrt{-g}\,C_{\mu\nu\rho\sigma}C^{\mu\nu\rho\sigma}$. Writing
$\delta g_{\mu\nu}=-\Lie_\zeta g_{\mu\nu}+2\varphi g_{\mu\nu}$ and
$g_{\mu\nu}=\eta_{\mu\nu}+h_{\mu\nu}$, one obtains
$\delta\eta_{\mu\nu}=-2\sigma\eta_{\mu\nu}+2\varphi\eta_{\mu\nu}$, so that
preservation of the chosen background requires
\begin{equation}
 \varphi=\sigma ,
\label{eq:phisigma}
\end{equation}
for which the background contribution~$2\sigma\eta_{\mu\nu}$ cancels, leaving
$2\sigma h_{\mu\nu}$ in the field variation. This derivation does not require
us subsequently to treat the background as dynamical; throughout the analysis,
$h_{\mu\nu}$ is a field on fixed Minkowski spacetime.

It is important to distinguish the function~$\sigma$, determined by the
spacetime generator~$\zeta^\mu$ and multiplying~$h_{\mu\nu}$, from the arbitrary
internal Weyl parameter~$\omega$, which multiplies~$\eta_{\mu\nu}$ in
$\delta_\omega h_{\mu\nu}=2\omega\eta_{\mu\nu}$. The former is required for
conformal invariance of the action; the latter remains a freely specifiable
gauge parameter. In particular, the weight term differs in this respect from
the optional Bessel-Hagen choices~$\alpha=A_\nu\zeta^\nu$ in electromagnetism
and~$\xi_\mu=h_{\mu\nu}\zeta^\nu$ in Fierz-Pauli, which supplement an existing
symmetry by a gauge variation. For the Poincar\'e generators considered in
Ref.~\cite{HobsonBarkerLasenbyFP}, $\zeta^\mu$ is a Killing vector and
$\sigma=0$, so no weight term occurs.

\section{The conformal Bessel-Hagen current and its class}
\label{sec:class}

\subsection{The general form variation and the pure-gauge current}

The internal gauge variations~\cref{eq:gauges} may be added to the conformal
variation~\cref{eq:Qzeta}, giving the general Bessel-Hagen form variation
\begin{equation}
 \Delta^{\zeta,\xi,\omega}_{\mu\nu}
 =\Delta^\zeta_{\mu\nu}+D_\xi h_{\mu\nu}+2\omega\eta_{\mu\nu}.
\label{eq:Qgen}
\end{equation}
To determine the effect of the compensating parameters on the current, consider
first a pure gauge variation~$G_{\mu\nu}=D_\xi h_{\mu\nu}+2\omega\eta_{\mu\nu}$.
Since~$C^{(1)}[G]=0$, the density is pointwise invariant,
$\delta_G\mathcal{L}_{\rm C}=0$. Using the Noether
identities~\cref{eq:bachidentities} in the variational
identity~\cref{eq:var}, one obtains
\begin{equation}
\begin{aligned}
 \mathcal{B}^{\mu\nu}D_\xi h_{\mu\nu}&=\partial_\mu(2\mathcal{B}^{\mu\nu}\xi_\nu), \\
 \mathcal{B}^{\mu\nu}(2\omega\eta_{\mu\nu})&=2\omega\,\mathcal{B}^\mu{}_\mu=0 .
\end{aligned}
\label{eq:puregaugevar}
\end{equation}
The algebraic Poincar\'e lemma then allows the associated current to be written
as a field-equation term and a superpotential,
\begin{equation}
 \Theta^\alpha(G)
 =-2\mathcal{B}^{\alpha\nu}\xi_\nu
 +\partial_\beta S^{[\alpha\beta]}_{\xi,\omega},
\label{eq:puregauge}
\end{equation}
so that it is trivial as a conserved current. The Weyl parameter~$\omega$
contributes only to the superpotential, since~$\mathcal{B}^\mu{}_\mu\equiv0$.

It is useful to compare this result with the Fierz-Pauli
relation~\cite{HobsonBarkerLasenbyFP}
$J^\mu_{\zeta,\xi}=J^\mu_{\zeta,0}-2\mathcal{E}^{\mu\nu}\xi_\nu$, where
$\mathcal{E}^{\mu\nu}$ is the linearised Einstein tensor.\footnote{This
Fierz-Pauli relation holds off shell for the natural representative of the
inexact gauge boundary term~$K^\mu_\xi$. For a different representative,
$K^\mu_\xi\mapsto K^\mu_\xi+\partial_\nu Y^{[\mu\nu]}_\xi$, it becomes
$J^\mu_{\zeta,\xi}=J^\mu_{\zeta,0}-2\mathcal{E}^{\mu\nu}\xi_\nu
-\partial_\nu(Y^{[\mu\nu]}_\xi-Y^{[\mu\nu]}_0)$.
The additional superpotential is trivial in the Barnich--Brandt--Henneaux
sense, and is analogous to the term in~\cref{eq:puregauge}.}
In Fierz-Pauli theory no choice of the compensating parameter gives a strictly
gauge-invariant local tensor, although the current class is gauge invariant.
There the field-equation contribution was obtained from an inexact gauge
boundary term, since the density is invariant only up to a divergence. In the
present theory the density is strictly gauge invariant, and the analogous
term~$-2\mathcal{B}^{\alpha\nu}\xi_\nu$ instead follows from the Ostrogradski
symplectic potential. Both expressions are consequences of Noether's second
identity, namely~$\partial_\mu\mathcal{E}^{\mu\nu}\equiv0$ and
$\partial_\mu\mathcal{B}^{\mu\nu}\equiv0$, respectively. Thus, although the
inexact boundary term provided the Noetherian explanation in the Fierz-Pauli
calculation, the freedom to add a field-equation term is a more general
property of the variational construction.

\subsection{The current and the~$(\xi,\omega)$-family}

For the conformal variation~\cref{eq:Qzeta}, the flat-space density obeys
$\delta_\zeta\mathcal{L}_{\rm C}=-\partial_\alpha
(\zeta^\alpha\mathcal{L}_{\rm C})$, as established in~\cref{sec:background}.
The Noether current for the combined variation~\cref{eq:Qgen} is consequently
\begin{equation}
 J^\alpha_{\zeta,\xi,\omega}
 =\Theta^\alpha(\Delta^{\zeta,\xi,\omega})+\zeta^\alpha\mathcal{L}_{\rm C}.
\label{eq:Jdef}
\end{equation}
Linearity of~$\Theta^\alpha$, together with~\cref{eq:puregauge}, then gives
\begin{equation}
 J^\alpha_{\zeta,\xi,\omega}
 =J^\alpha_{\zeta,0,0}
 -2\mathcal{B}^{\alpha\nu}\xi_\nu
 +\partial_\beta S^{[\alpha\beta]}_{\xi,\omega}
\label{eq:Jfamily}
\end{equation}
where~$J^\alpha_{\zeta,0,0}=\Theta^\alpha(\Delta^\zeta)+\zeta^\alpha\mathcal{L}_{\rm C}$
is the representative obtained without an internal compensation. It follows
that every choice of~$\xi$ and~$\omega$ gives the same conserved-current class,
since their contributions consist of a Bach-equation term and a
superpotential.\footnote{Both contributions are trivial in the
Barnich--Brandt--Henneaux sense, but their off-shell divergences differ. For any
antisymmetric~$U^{[\alpha\beta]}$, one has
$\partial_\alpha\partial_\beta U^{[\alpha\beta]}\equiv0$, whereas
$\partial_\alpha(-2\mathcal{B}^{\alpha\nu}\xi_\nu)
=-\mathcal{B}^{\alpha\nu}D_\xi h_{\alpha\nu}$ by~\cref{eq:bachidentities}.
The latter is the Euler--Lagrange contraction and vanishes on the Bach shell.
In the displayed representative, the parameter dependence of the off-shell
divergence is therefore carried by the Bach term; the remaining dependence,
including that of~$S^{[\alpha\beta]}_{\xi,\omega}$, may be included in an
identically conserved superpotential. The separation of weakly vanishing and
superpotential terms is not itself completely canonical.}

As in Fierz-Pauli theory, choosing~$\xi_\mu=h_{\mu\nu}\zeta^\nu$ gives a form
variation expressed in terms of the linearised connection,
$\Delta^{\zeta,h\zeta,0}_{\mu\nu}=2\zeta^\rho\Gamma^{(1)}_{\rho\mu\nu}
+2\sigma h_{\mu\nu}$, where
$\Gamma^{(1)}_{\rho\mu\nu}=\half(\partial_\mu h_{\nu\rho}
+\partial_\nu h_{\mu\rho}-\partial_\rho h_{\mu\nu})$.
This choice does not change the current class. Although~\cref{eq:Jfamily} was
derived for arbitrary external functions~$\xi_\mu(x)$ and~$\omega(x)$,
field-dependent choices are obtained by substitution: the pure-gauge identity
uses only the off-shell transversality and tracelessness of the Bach tensor
and continues to hold for such choices. This is analogous to the dependence
of the Fierz-Pauli current on~$\xi$ through the transverse Einstein
tensor~\cite{HobsonBarkerLasenbyFP}. When a further independent gauge variation
is performed, however, the field dependence of the compensating parameters
must also be varied, as we now demonstrate. Neither the current-class result
nor the no-go theorem in~\cref{sec:nogo} depends on selecting a particular
compensation.

\subsection{Gauge invariance of the class for the full conformal group}

We now verify that the current class is gauge invariant for every conformal
generator. Consider an independent internal gauge transformation
$\delta_{\chi,\psi}h_{\mu\nu}=D_\chi h_{\mu\nu}+2\psi\eta_{\mu\nu}$.
To determine its action on the conformal form variation, one may use
$\Lie_\zeta(D_\chi h)_{\mu\nu}=D_{\Lie_\zeta\chi}h_{\mu\nu}
-2\chi_\rho\partial_\mu\partial_\nu\zeta^\rho$, together with the conformal
Killing relation~\cref{eq:ddckv}. A direct calculation gives
\begin{equation}
 \delta_{\chi,\psi}\Delta^\zeta_{\mu\nu}
 =D_\eta h_{\mu\nu}+2\kappa\,\eta_{\mu\nu},
\label{eq:dQ}
\end{equation}
with
\begin{equation}
 \eta_\mu=2\sigma\chi_\mu-(\Lie_\zeta\chi)_\mu,
 \qquad
 \kappa=-\chi^\rho\partial_\rho\sigma-\zeta^\rho\partial_\rho\psi .
\label{eq:etakappa}
\end{equation}
Thus the gauge variation of~$\Delta^\zeta$ is itself an internal gauge
variation. In particular, for special conformal transformations the terms
containing~$\partial_\mu\partial_\nu\zeta^\rho$ combine, using
\cref{eq:ddckv,eq:phisigma}, into the Weyl contribution
$\kappa\supset-\chi^\rho\partial_\rho\sigma$. The internal Weyl symmetry
therefore accommodates the additional terms that occur when the spacetime
generators are extended beyond the Poincar\'e group.

If the compensating parameters depend on the field, the same calculation gives
$\eta_\mu\to\bar\eta_\mu=\delta_{\chi,\psi}\xi_\mu+2\sigma\chi_\mu
-(\Lie_\zeta\chi)_\mu$ and a shift of~$\kappa$ by
$\delta_{\chi,\psi}\omega$. The latter affects only the superpotential, since
the pure Weyl current has no field-equation term when
$\mathcal{B}^\mu{}_\mu\equiv0$. Since the Bach tensor is also gauge invariant,
$\delta_{\chi,\psi}\mathcal{B}^{\mu\nu}=0$,
\cref{eq:Jfamily,eq:puregauge,eq:dQ} imply
\begin{equation}
 \delta_{\chi,\psi}J^\alpha_{\zeta,\xi,\omega}
 =-2\mathcal{B}^{\alpha\nu}\bar\eta_\nu
 +\partial_\beta U^{[\alpha\beta]}_{\zeta;\chi,\psi},
\label{eq:classvar}
\end{equation}
so that on shell
\begin{equation}
 \delta_{\chi,\psi}J^\alpha_{\zeta,\xi,\omega}\eom
 \partial_\beta U^{[\alpha\beta]}_{\zeta;\chi,\psi}.
\label{eq:classinv}
\end{equation}
This establishes gauge invariance of the conserved-current class for every
conformal generator, including dilations and special conformal transformations.
The latter introduce no additional obstruction because their second-derivative
terms are included in~$\kappa$. This extends the Fierz-Pauli current-class
result, but does not establish that all the currents are coordinate moments
of a single canonical rank-two tensor, as in electromagnetism. Such an
identification would require improvements involving spin, virial and
higher-derivative contributions, and is not needed for the class-invariance
argument.

\section{Non-existence of a local gauge-invariant representative}
\label{sec:nogo}

We now ask whether the translation current admits a strictly gauge-invariant
local energy-momentum tensor. Both transformations in~\cref{eq:gauges} are
treated as gauge redundancies of the pure linearised Weyl-squared theory, and
the representative is required to be invariant under each. Adopting the Einstein
gauge for a compensator does not by itself break the Weyl
symmetry~\cite{cgrotcurves}; likewise, boundary conditions fixing a conformal
representative leave compactly supported Weyl transformations admissible in
the interior. Explicit Weyl-breaking terms in the bulk action would change
the assumptions and require a separate analysis, while gauge fixing and
extensions involving additional fields raise distinct questions discussed
in~\cref{sec:discussion}.

For translations~$\zeta^\mu=a^\mu$, the representative without compensation
satisfies~$J^\alpha_{a,0,0}=-a^\rho t^\alpha{}_\rho$, where the canonical tensor
is
\begin{equation}
 t^\alpha{}_\rho
 =\Pi^{\alpha\beta|\mu\nu}\partial_\beta\partial_\rho h_{\mu\nu}
 -\big(\partial_\beta\Pi^{\alpha\beta|\mu\nu}\big)\partial_\rho h_{\mu\nu}
 -\delta^\alpha_\rho\mathcal{L}_{\rm C}.
\label{eq:tcanon}
\end{equation}
Although~$t^\alpha{}_\rho$ is conserved on the Bach shell, its gauge invariance
holds only at the level of the current class. We therefore consider whether
an improvement of the form
\begin{equation}
 T^\alpha{}_\rho=t^\alpha{}_\rho
 +\partial_\beta V^{[\alpha\beta]}{}_\rho
 +\mathcal{B}^{\alpha\nu}Y_{\nu\rho},
\label{eq:improve}
\end{equation}
can give a strictly gauge-invariant local tensor. Here strict invariance means
$\delta_\xi T^\alpha{}_\rho=\delta_\omega T^\alpha{}_\rho=0$ pointwise, without
using the field equations or identifying tensors that differ by a
superpotential. It is consequently stronger than the invariance of the
current class established in~\cref{eq:classinv}.

The improvements in~\cref{eq:improve} preserve on-shell conservation, but we
admit only those for which the resulting tensor remains quadratic in
$h_{\mu\nu}$ and has mass dimension four, as appropriate to the quadratic
action~\cref{eq:action}, whose density has~$[\mathcal{L}_{\rm C}]=4$.
More precisely, let~$\mathcal{A}$ consist of tensors~$T_{\mu\nu}$ that are
(i) local, Poincar\'e-covariant and polynomial in~$h_{\mu\nu}$ and finitely many
of its derivatives, with no explicit coordinate dependence and only constant
dimensionless coefficients, so that no external mass or length scale is
introduced; (ii) symmetric and of rank two; (iii) of mass dimension four and
quadratic in~$h_{\mu\nu}$; (iv) strictly invariant under both gauge symmetries;
and (v) conserved on the Bach shell. We will show that~$\mathcal{A}=\{0\}$.

The restrictions on coefficients and coordinate dependence are needed for the
dimension argument below: expressions such as~$\ell^2(\partial C^{(1)})^2$ or
$x^2(\partial C^{(1)})^2$ do not belong to the allowed class of
translation-covariant stress tensors for the scale-free theory. These
restrictions are assumptions on the resulting improved tensor and are not
properties of an arbitrary superpotential or field-equation term. We also
require symmetry, as appropriate to a local stress-energy representative;
an antisymmetric part of a canonical translation tensor must be removed by
relocalisation before that interpretation is made. With these assumptions,
the proof consists of four steps.

\paragraph*{Step 1: dependence on the gauge-invariant curvature.}
A standard result of the local cohomology of linearised gravity is that local
diffeomorphism invariants of~$h_{\mu\nu}$ can be expressed in terms of
$R^{(1)}_{\mu\nu\rho\sigma}$ and its derivatives; in particular, first
derivatives of~$h_{\mu\nu}$ can be set to zero at a point by a gauge
transformation~\cite{DeserHenneaux1995,HobsonBarkerLasenbyFP}. Requiring invariance
also under the arbitrary internal Weyl variation
$\delta_\omega h_{\mu\nu}=2\omega\eta_{\mu\nu}$ restricts this dependence to
the Weyl curvature and its derivatives. The relevant transformation is distinct
from the fixed weight contribution~$\sigma$ in~\cref{sec:background}; it gives
\begin{equation}
\begin{aligned}
 \delta_\omega R^{(1)}_{\mu\nu}&=-2\partial_\mu\partial_\nu\omega
 -\eta_{\mu\nu}\Box\omega\neq0, \\
 \delta_\omega C^{(1)}_{\mu\nu\rho\sigma}&=0 .
\end{aligned}
\label{eq:domegaric}
\end{equation}
Thus independent Ricci-dependent contributions are excluded. The allowed
local invariant tensors can therefore be constructed from~$C^{(1)}$ and its
derivatives.

\paragraph*{Step 2: dimension and field degree.}
In the normalisation~$S_{\rm C}=\half\int C^{(1)2}$, the field is dimensionless,
$[h]=0$, and~$[C^{(1)}]=2$. This differs from the canonically normalised
Fierz-Pauli field, which has mass dimension one. A tensor of dimension four
that is quadratic in~$h$ and constructed from~$C^{(1)}$ can consequently contain
two Weyl factors but no further derivatives.

For contractions using the metric, six of the eight indices must be contracted
and two left free. Contracting a pair within one Weyl factor gives a trace and
hence zero; likewise, leaving both free indices on one factor forces a trace
on the other. Apart from the scalar contraction multiplied by~$\eta_{\mu\nu}$,
one must therefore leave one free index on each factor. The parity-even
symmetric rank-two candidates are
\begin{equation}
 C^{(1)}_{\mu\alpha\beta\gamma}C^{(1)}_\nu{}^{\alpha\beta\gamma}
 \quad\text{and}\quad
 \eta_{\mu\nu}\,C^{(1)}_{\rho\sigma\lambda\tau}C^{(1)\rho\sigma\lambda\tau},
\label{eq:basis}
\end{equation}
while the parity-odd sector is obtained by replacing one Weyl factor with its dual,
${}^*\!C^{(1)}_{\mu\nu\rho\sigma}=\half\epsilon_{\mu\nu}{}^{\alpha\beta}
C^{(1)}_{\alpha\beta\rho\sigma}$, giving
$C^{(1)}_{\mu\alpha\beta\gamma}{}^*\!C^{(1)}_\nu{}^{\alpha\beta\gamma}$ and
$\eta_{\mu\nu}C^{(1)}{}^*\!C^{(1)}$.
The pair symmetries and first Bianchi identity reduce other index arrangements
to these four structures.

\paragraph*{Step 3: four-dimensional algebraic identities.}
In four dimensions the parity-even contraction satisfies
\begin{equation}
 C^{(1)}_{\mu\alpha\beta\gamma}C^{(1)}_\nu{}^{\alpha\beta\gamma}
 =\tfrac14\eta_{\mu\nu}\,C^{(1)}_{\rho\sigma\lambda\tau}C^{(1)\rho\sigma\lambda\tau}
\label{eq:4Did}
\end{equation}
which follows by tracing the identically vanishing antisymmetrisation over
five indices. A related identity accounts for the topological character of
the four-dimensional Gauss--Bonnet term. The parity-odd contraction similarly
obeys
\begin{equation}
 C^{(1)}_{\mu\alpha\beta\gamma}{}^*\!C^{(1)}_\nu{}^{\alpha\beta\gamma}
 =\tfrac14\eta_{\mu\nu}\,C^{(1)}{}^*\!C^{(1)} .
\label{eq:4Didodd}
\end{equation}

A useful derivation of both identities is obtained in two-spinor
notation~\cite{PenroseRindler}.\footnote{We adopt the two-spinor conventions of
Ref.~\cite{PenroseRindler}: spinor indices are raised and lowered with the
antisymmetric~$\epsilon_{AB}=-\epsilon_{BA}$ by~$\psi^A=\epsilon^{AB}\psi_B$ and
$\psi_B=\psi^A\epsilon_{AB}$ (so that~$\epsilon^{AB}\epsilon_{CB}=\delta^A{}_C$),
and the dual is as in the main text. The signs of the intermediate relation
\cref{eq:spinorcontr} and of the~$\mp i$ factors below depend on these choices;
the final identities~\cref{eq:4Did} and~\cref{eq:4Didodd}, fixed by their own
traces, are convention-independent.} Writing a vector index as a pair~$a=AA'$, the
linearised Weyl tensor decomposes into self-dual and anti-self-dual parts,
\begin{equation}
 C^{(1)}_{abcd}=\Psi_{ABCD}\,\epsilon_{A'B'}\epsilon_{C'D'}
 +\bar\Psi_{A'B'C'D'}\,\epsilon_{AB}\epsilon_{CD},
\label{eq:spinordecomp}
\end{equation}
where~$\Psi_{ABCD}$ is totally symmetric and dualisation multiplies the two
parts by~$\mp i$. In the contraction~$C^{(1)}_{m\,acd}C^{(1)}_n{}^{acd}$, with
$m=MM'$ and~$n=NN'$, the mixed self-dual and anti-self-dual terms vanish, since
contraction of a symmetric spinor with~$\epsilon$ gives
$\Psi_{MACD}\epsilon^{CD}=0$. For the remaining terms, the contraction within
one chirality is antisymmetric in its free indices:
\begin{equation}
 \Psi_{MACD}\,\Psi_N{}^{ACD}=\tfrac12\,\epsilon_{MN}\,\Psi_{PQRS}\Psi^{PQRS},
\label{eq:spinorcontr}
\end{equation}
where interchanging~$M$ and~$N$ reverses the sign because there are three
$\epsilon$-contractions. Thus
$C^{(1)}_{m\,acd}C^{(1)}_n{}^{acd}\propto
\epsilon_{MN}\epsilon_{M'N'}=\eta_{mn}$, and taking the trace fixes the
coefficient in~\cref{eq:4Did}. Inserting a dual changes only the factors
$\mp i$ in the two chiral sectors, giving~\cref{eq:4Didodd} in the same way.
These reductions depend on the spacetime dimension: for~$n>4$, the trace-free
part of~$C^{(1)}_{\mu\alpha\beta\gamma}C^{(1)}_\nu{}^{\alpha\beta\gamma}$ need
not vanish, and this argument does not establish an analogous obstruction.

Every candidate satisfying the preceding assumptions is therefore of the form
\begin{equation}
\begin{aligned}
 T_{\mu\nu}&=\eta_{\mu\nu}\big(c_1\,C^{(1)2}+c_2\,C^{(1)}{}^*\!C^{(1)}\big), \\
 c_1,c_2&=\text{const}.
\end{aligned}
\label{eq:Ttrace}
\end{equation}
In particular, no trace-free part remains.

\paragraph*{Step 4: conservation on the Bach shell.}
For~\cref{eq:Ttrace} to be conserved on every Bach solution, one would require
$\partial^\mu T_{\mu\nu}=\partial_\nu
(c_1C^{(1)2}+c_2C^{(1)}{}^*\!C^{(1)})=0$ on shell.
The two coefficients can be constrained separately using Ricci-flat
configurations, which are also Bach-flat. In the vacuum region~$r>0$, the
linearised Schwarzschild field has~$C^{(1)2}\propto M^2/r^6$ and vanishing
Pontryagin density. Conservation on this configuration therefore requires
$c_1=0$, but places no restriction on~$c_2$.

To constrain the latter coefficient, consider the linearised Lense--Thirring
field with mass~$M$ and angular momentum~$J$, which is likewise Ricci-flat for
$r>0$. Its Pontryagin density contains the non-constant mass--spin term
$C^{(1)}{}^*\!C^{(1)}\propto MJ\cos\theta/r^7$, while the~$M^2$ and~$J^2$
contributions vanish. Only the non-constancy of this leading asymptotic term
is needed: its overall coefficient depends on conventions, and higher spin
multipoles of the full Kerr solution do not enter the argument. Hence
$c_2=0$ also, and the only conserved tensor in the allowed family is zero.

\medskip
\noindent
It follows that~$\mathcal{A}=\{0\}$. To exclude a strict representative of the
translation current, one must also establish that its conserved-current class
is nonzero. Rigid translations act on the gauge-invariant curvature by
$\delta_a C^{(1)}_{\mu\nu\rho\sigma}
=-a^\lambda\partial_\lambda C^{(1)}_{\mu\nu\rho\sigma}$, and on generic Bach
solutions this action is neither an internal gauge transformation nor an
on-shell trivial symmetry. Under the standard regularity assumptions of the
inverse Noether theorem~\cite{BarnichBrandtHenneaux2000}, the corresponding
current class is therefore nonzero. It cannot be represented by the only
tensor in~$\mathcal{A}$, and consequently no improvement of the canonical
tensor~\cref{eq:tcanon} satisfying the stated restrictions yields a strictly
gauge-invariant local representative. In this sense, the trace-free part of
the conserved translation tensor cannot be made gauge invariant.

The restriction to rank-two tensors is relevant here. On the Einstein subset
of the Bach solution space there is a familiar gauge-invariant rank-four
quantity, the linearised Bel--Robinson
tensor~\cite{Bel1958,Bel1962,Senovilla2000},
\begin{equation}
 T_{\alpha\beta\gamma\delta}
 =C^{(1)}_{\alpha\mu\gamma\nu}C^{(1)}{}_{\beta}{}^{\mu}{}_{\delta}{}^{\nu}
 +{}^*\!C^{(1)}_{\alpha\mu\gamma\nu}{}^*\!C^{(1)}{}_{\beta}{}^{\mu}{}_{\delta}{}^{\nu},
\label{eq:belrobinson}
\end{equation}
which is totally symmetric, trace-free and invariant under both gauge
symmetries because it is constructed from~$C^{(1)}$. The rank-two
identity~\cref{eq:4Did} does not make this rank-four expression vanish. Its
divergence is given, in our conventions, by a contraction of~$C^{(1)}$ with
the linearised Cotton tensor
$K_{\mu\gamma\nu}\equiv\partial^\alpha C^{(1)}_{\alpha\mu\gamma\nu}$.
By the contracted Bianchi identity, the latter is a derivative of the
linearised Ricci tensor and vanishes when~$R^{(1)}_{\mu\nu}=0$. The
Bel--Robinson tensor is thus conserved on these linearised-vacuum Einstein
configurations. The Bach equation~$\mathcal{B}^{\mu\nu}=0$ is weaker, however,
and does not require the Ricci or Cotton tensor to vanish, so this argument
does not establish conservation on generic non-Einstein Bach modes.

The Bel--Robinson tensor is not obtained here from the Bessel-Hagen current;
it is exhibited as a gauge-invariant super-energy tensor on the specified
subset of solutions. Its algebraic construction nevertheless provides a
useful comparison: the rank-four Noether momentum is a double-symmetric
projection of the Weyl tensor, from which a quadratic, totally symmetric,
trace-free rank-four expression can be formed, just as the Maxwell tensor is
quadratic in~$\Pi^{\mu|\sigma}=-F^{\mu\sigma}$. Contracting the Weyl factors
to rank two instead removes the trace-free part by~\cref{eq:4Did}, and the
remaining trace fails the conservation test above. This parallels the
Bel--Robinson observation in the Fierz-Pauli
companion~\cite{HobsonBarkerLasenbyFP}, where the corresponding expression is
quadratic in~$R^{(1)}$. Whether a rank-four tensor conserved on the full Bach
shell exists remains a separate question.

\section{Discussion}
\label{sec:discussion}

The comparison with electromagnetism, Fierz-Pauli and linearised Gauss--Bonnet
gravity is summarised in~\cref{tab:fsorder}. Two aspects of the construction
are relevant: the derivative order of the invariant field strength and its
field equation, and the algebraic identities satisfied by the quadratic
rank-two expressions. In electromagnetism, the first-order field strength
$F\sim\partial A$ admits a nonzero quadratic Maxwell tensor, whose divergence
vanishes using~$\partial_\mu F^{\mu\nu}=0$. For Fierz-Pauli, the obstruction
already concerns expressions of order~$\partial h\,\partial h$, since the
action contains first derivatives while the first gauge-invariant field
strength contains two. The present result shows that replacing this action
by one built from the Weyl curvature does not suffice to obtain a strictly
gauge-invariant local stress tensor.

\begin{table*}
\caption{\label{tab:fsorder}Comparison of local gauge-invariant energy-momentum
constructions in four dimensions. For the Weyl-squared theory the negative
result refers to the admissible class~$\mathcal{A}$ defined in~\cref{sec:nogo}.
The Gauss--Bonnet entry concerns its proposed Riemann-quadratic tensor, which
vanishes identically; the listed symmetry specifies the test on that candidate
and does not classify all symmetries of the topological action.}

\begin{ruledtabular}
\begin{tabular}{lccc}
theory & action quadratic in & gauge symmetry & strict local~$T_{\mu\nu}$\\
\hline
electromagnetism &~$F_{\mu\nu}\sim\partial A$ &~$U(1)$ & yes\\
linearised Gauss--Bonnet &~$R^{(1)}\sim\partial\partial h$ & diffeos (candidate test; action topological) & no\\
linearised~$C^2$ &~$C^{(1)}\sim\partial\partial h$ & diffeomorphisms~$+$ Weyl & no\\
Fierz-Pauli &~$\partial h$ (connection-like) & diffeomorphisms & no\\
\end{tabular}
\end{ruledtabular}
\end{table*}
The Gauss--Bonnet example is useful because its proposed rank-two tensor can
be tested under linearised diffeomorphisms without imposing strict Weyl
invariance on the candidate. Riemann-dependent expressions are then allowed,
but the tensor constructed by Baker and Kuzmin~\cite{BakerKuzmin2019},
\begin{equation}
\begin{split}
 T^{\mu\nu}_{\rm GB}={}&-R^{\mu\rho\lambda\sigma}R^\nu{}_{\rho\lambda\sigma}
 +2R_{\rho\sigma}R^{\mu\rho\nu\sigma}\\
 &+2R^{\mu\lambda}R^\nu{}_\lambda-RR^{\mu\nu}
 +\tfrac14\eta^{\mu\nu}\mathcal{G},
\end{split}
 \label{eq:bkgb}
\end{equation}
where~$\mathcal{G}$ is the Gauss--Bonnet scalar, equals~$-\tfrac12$ times the
Lanczos tensor and therefore vanishes identically in four dimensions. This is
a consequence of the topological character of the Gauss--Bonnet term.
Although Ref.~\cite{BakerKuzmin2021} argues that the action contributes to this
tensor despite its vanishing field equations, the displayed contribution is
itself zero. Thus its algebraic similarity to the Maxwell expression does not
yield a nonzero local energy-momentum tensor.

The two curvature-quadratic examples fail in different ways. For Weyl-squared
gravity, invariance under both gauge symmetries restricts the candidates to
Weyl-dependent expressions; the four-dimensional identities leave only pure
traces, whose conservation is not implied by the Bach equation. For
Gauss--Bonnet, the Riemann-dependent candidate vanishes by the four-dimensional
Lovelock (Lanczos) identity. In each case a dimension-dependent curvature
identity prevents the proposed electromagnetic correspondence, but in the
Weyl-squared case the restriction to the Weyl curvature is also an essential
part of the proof. Conformal invariance alone does not explain the difference
from electromagnetism, which is itself conformally invariant.

The internal Weyl symmetry also enters the current-class calculation in a
different, compatible way. Its arbitrary parameter allows the additional
terms from special conformal transformations to be included in the pure-gauge
variation~\cref{eq:dQ}; requiring invariance of a local representative under
the same symmetry restricts its dependence to~$C^{(1)}$ and its derivatives.
These are two consequences of the same gauge structure, with the algebraic
identity needed in addition for the non-existence result. The Bessel-Hagen
construction accordingly provides a gauge-invariant current class for every
conformal generator, while its translation class has no strict local
representative within~$\mathcal{A}$.

The electromagnetic example suggests investigating theories with independent
first-order gauge-covariant field strengths. In Weyl gauge theory of
gravity~\cite{eWGTpaper,CGTpaper}, the dilation curvature
$H_{\mu\nu}=\partial_\mu B_\nu-\partial_\nu B_\mu$ of the scale
connection~$B_\mu$ has the form of the electromagnetic field strength. The
manifestly homogeneous gauge-covariant currents obtained for this
theory~\cite{HobsonLasenbyBarker2024} therefore motivate a Bessel-Hagen analysis
of this internal field-strength sector, which is left to future work. Such a
field strength avoids the specific derivative-counting and curvature-identity
arguments used here, but its presence alone does not establish a local
representative with the required conservation and gauge properties.

In particular, our theorem concerns a metric formulation and does not cover
theories with an independent connection and coframe. Two companion papers
address related questions. The first~\cite{HobsonBarkerLasenbyEMT} establishes a
no-go result for Fierz-Pauli and the dynamically regular curvature-squared
metric spin-2 theories under specified locality, order and regularity
assumptions. It also identifies the fact that spacetime translations are
reducibility parameters of their gauge symmetry, without asserting that this
property alone implies the no-go result. Independent-connection formulations
lie outside that analysis.

The second companion~\cite{HobsonBarkerLasenbyT4} introduces an internal
St\"uckelberg~$T(4)$ redundancy in Weyl gauge theory, acting on the
St\"uckelberg pair while leaving the dressed coframe unchanged. It distinguishes
the total coframe Euler derivative, which vanishes on shell, from the explicit
gravitational self-source in an excitation/source equation. Soldering the latter
with the coframe yields a density invariant under the internal gauge factors,
with dilation invariance understood in the scale-invariant formulation; a frame
index therefore presents no obstruction to that construction. Two sufficient
charge constructions are given, neither of which is shown to be necessary.
The density depends on the gravitational/matter action split and the boundary
representative, however, and its ordinary conservation has not been established,
even on a flat homogeneous branch. Nor has it been identified with the external
rigid-translation current class. It consequently exhibits a local internal
source outside the domain of our theorem, while leaving open the existence of
a Maxwell-type representative of the external translation class.

It is also useful to distinguish our result from the energy-momentum
pseudo-tensor recently obtained by Faria~\cite{Faria2025} through a generalised
Bessel-Hagen method and used to calculate the energy of plane gravitational
waves. That work treats several conformally invariant theories, rather than
only the pure linearised~$h_{\mu\nu}$ sector, and its representative-dependent
object lies outside~$\mathcal{A}$. For the pure Weyl-squared sector our theorem
excludes a nonzero dimension-four quadratic tensor strictly invariant under
both gauge symmetries. Selecting a conformal representative may allow a
frame-dependent local expression, but does not by itself make that expression
gauge invariant. The canonical tensor~\cref{eq:tcanon} is itself the
Ostrogradski-type object discussed in the Introduction; the non-existence
argument does not depend on the sign of its energy.

Finally, the physical interpretation depends on the field content and bulk
gauge symmetries. In the theory considered here, the gauge-invariant
conserved-current class remains available, and on the Ricci-flat Einstein
subset of Bach solutions one also has the conserved Bel--Robinson
super-energy tensor~\cref{eq:belrobinson}. Explicit Weyl-breaking terms in the
bulk action, a gauge-invariant extension with additional fields, a gauge
fixing and a restriction at the boundary are distinct modifications. Only the
first removes the local Weyl redundancy in the interior. In particular,
fixing the conformal representative at the boundary still permits compactly
supported Weyl transformations in the bulk and does not admit a strict
Ricci-dependent local tensor there.

If genuinely physical scale-setting structure is introduced in the bulk, a
local expression tied to that structure may acquire physical meaning, but it
requires a separate construction and conservation test in the modified theory.
Whether it is a spacetime tensor invariant under the remaining diffeomorphism
symmetry cannot be inferred from the present calculation. This possibility
must be distinguished from setting a compensator to a constant in the Einstein
gauge, which alone does not break local scale
invariance~\cite{cgrotcurves}. Likewise, whether the conformal-gravity account
of galactic rotation curves~\cite{MannheimKazanas1989}, critically examined in
Ref.~\cite{cgrotcurves}, includes such physical structure is a question about
that theory. Neither the introduction of a scale nor the choice of a boundary
conformal representative alone settles the existence of an appropriate local
energy-momentum tensor.

\section{Conclusions}

We have applied the Bessel-Hagen construction to linearised Weyl-squared
gravity to determine whether replacing the first-derivative Fierz-Pauli action
by an action quadratic in a gauge-invariant curvature permits a local
energy-momentum tensor analogous to that of electromagnetism. The Noether
momentum is a symmetric projection of the linearised Weyl tensor and is
therefore gauge invariant. Moreover, the construction extends to the conformal
group, acting on the field on a fixed Minkowski background, provided the field
variation includes the weight term~$2\sigma h_{\mu\nu}$. The additional terms
arising for special conformal transformations are accommodated by an internal
Weyl gauge variation, and each generator defines a gauge-invariant
conserved-current class. The classes associated with different generators,
such as translations and rotations, are distinct; the invariance proved here
does not identify them with one another.

Despite these properties, the translation class admits no nonzero strictly
gauge-invariant local representative in the admissible family~$\mathcal{A}$.
For local, Poincar\'e-covariant, polynomial, symmetric rank-two tensors of mass
dimension four, quadratic in~$h_{\mu\nu}$, with no explicit coordinate
dependence or external scale, the two gauge symmetries restrict the candidates
to quadratic Weyl expressions without additional derivatives. Four-dimensional
identities reduce these to pure traces, and conservation on the Bach shell
requires both their parity-even and parity-odd coefficients to vanish. Thus
$\mathcal{A}=\{0\}$, whereas the translation current class is nonzero. On the
Ricci-flat Einstein subset of the Bach shell, the rank-four Bel--Robinson
tensor provides a conserved gauge-invariant super-energy, but no corresponding
conservation result for generic Bach modes has been established here.

The Gauss--Bonnet comparison gives a different instance in which an apparent
curvature-quadratic analogue of the Maxwell tensor fails: its proposed
energy-momentum tensor is the identically vanishing four-dimensional Lanczos
tensor. These examples show the importance of the dimension-dependent
curvature identities, in addition to the gauge restrictions on the candidates.
They motivate the investigation of independent first-order gauge-covariant
field strengths, such as the dilation curvature in Weyl gauge theory, although
such a field strength is not by itself sufficient to establish a Maxwell-type
local representative. The related linearised metric criterion and the separate
internal-translation construction are developed in companion
papers~\cite{HobsonBarkerLasenbyEMT,HobsonBarkerLasenbyT4}; the Bessel-Hagen
analysis of the dilation sector remains to be performed.

\begin{acknowledgments}
%This work was supported by the research environment and infrastructure of the Handley Lab at the University of Cambridge.
W.~B. is grateful for the support of Girton College, Cambridge, Marie
Sk\l{}odowska-Curie Actions, and the hospitality of the Helsinki
Institute of Physics. Co-funded by the European Union (Physics for Future -- Grant Agreement No. 101081515). Views and opinions expressed are however those of the author(s) only and do not necessarily reflect those of the European Union or European Research Executive Agency. Neither the European Union nor the granting authority can be held responsible for them.
\end{acknowledgments}

\end{document}